\documentclass{aastex62}

\usepackage{amsfonts,times}
\usepackage{mathrsfs}
\usepackage{rotating}

\def\cblue{\color{black}}

\def\sunm{M_{\odot}}

\newcommand{\BHm}{M_{\bullet}}
\newcommand{\bhm}{m_{\bullet}}
\newcommand{\bhmp}{m_{\rm p}}
\newcommand{\bhms}{m_{\rm s}}

\newcommand{\calA}{{\cal A}}
\newcommand{\calJ}{{\cal J}}

\newcommand{\calT}{{\cal T}_{\rm tid}}

\newcommand{\dotMBon}{\dot{M}_{\rm Bon}}
\newcommand{\MBon}{M_{\rm Bon}}
\newcommand{\RBon}{R_{\rm Bon}}
\newcommand{\Rg}{R_{\rm g}}

\newcommand{\mathdotM}{\dot{\mathscr{M}}}

\newcommand\kms{\rm km\,s^{-1}}

\begin{document}


\title{\large\bf Accretion-modified Stars in Accretion Disks of Active Galactic Nuclei: Gravitational 
Wave Bursts \\  \vglue 0.1cm and Electromagnetic Counterparts from Merging Stellar Black Hole Binaries}

\author[0000-0001-9449-9268]{Jian-Min Wang}
\affil{Key Laboratory for Particle Astrophysics, Institute of High Energy Physics,
Chinese Academy of Sciences, 19B Yuquan Road, Beijing 100049, China}
\affil{School of Astronomy and Space Sciences, University of Chinese Academy of Sciences, 
19A Yuquan road, Beijing 100049, China}
\affil{National Astronomical Observatory of China, 20A Datun Road, Beijing 100020, China}

\author{Jun-Rong Liu}
\affil{Key Laboratory for Particle Astrophysics, Institute of High Energy Physics,
Chinese Academy of Sciences, 19B Yuquan Road, Beijing 100049, China}
\affil{School of Astronomy and Space Sciences, University of Chinese Academy of Sciences, 
19A Yuquan road, Beijing 100049, China}

\author{Luis C. Ho}
\affil{Kavli Institute for Astronomy and Astrophysics, Peking University, Beijing 100871, China}
\affil{Department of Astronomy, School of Physics, Peking University, Beijing 100871, China}

\author{Yan-Rong Li}
\affil{Key Laboratory for Particle Astrophysics, Institute of High Energy Physics,
Chinese Academy of Sciences, 19B Yuquan Road, Beijing 100049, China}

\author{Pu Du}
\affil{Key Laboratory for Particle Astrophysics, Institute of High Energy Physics,
Chinese Academy of Sciences, 19B Yuquan Road, Beijing 100049, China}


\begin{abstract}
The recent advanced LIGO/Virgo detections of gravitational waves (GWs) from stellar binary black hole 
(BBH) mergers, in particular GW190521, which is potentially associated with a quasar, have stimulated 
renewed interest in active galactic nuclei (AGNs) as factories of merging BBHs. Compact objects 
evolving from massive stars are unavoidably enshrouded by a massive envelope to form accretion-modified 
stars (AMSs) in the dense gaseous environment of a supermassive black hole (SMBH) accretion disk. We 
show that most AMSs form binaries due to gravitational interaction with each other during radial 
migration in the SMBH disk, forming BBHs inside the AMS. When a BBH is born, its orbit is initially 
governed by the tidal torque of the SMBH. Bondi accretion onto BBH at a hyper-Eddington rate naturally
develops and then controls the evolution of its orbits. We find that Bondi accretion leads to efficient
removal of orbital angular momentum of the binary, whose final merger produces a GW burst. Meanwhile, 
the Blandford-Znajek mechanism pumps the spin energy of the merged BH to produce an electromagnetic
counterpart (EMC). Moreover, hyper-Eddington accretion onto the BBH develops powerful outflows and 
triggers a Bondi explosion, which manifests itself as a EMC of the GW burst, depending on the viscosity 
of the accretion flow. Thermal emission from Bondi sphere appears as one of EMCs. BBHs radiate GWs with
frequencies $\sim 10^{2}\,$Hz, which  are accessible to LIGO.
\end{abstract}
\subjectheadings{Active galactic nuclei (16); Galaxy accretion disks (562); Supermassive black holes (1663)}

\section{Introduction}
The outer parts of the accretion disk of supermassive black holes (SMBHs) in active galactic nuclei 
(AGNs) host many poorly understood, complicated processes.  Star formation is unavoidable in these 
regions because of 
self-gravity \citep{Paczynski1977,Kolykhalov1980,Shlosman1989,Collin1999,Goodman2003,Goodman2004,Collin2008}, 
producing compact stellar remnants from the rapid evolution of massive 
stars \citep{Artymowicz1993,Cheng1999,Cantiello2020,Moranchel2021,Wang2021,Grishin2021}.  Stellar 
evolution rapidly 
releases metals into the outer parts of the self-gravitating (SG) disk \citep{Wang2010,Wang2011,Wang2012}, 
offering an explanation for the super-solar metallicities observed in AGNs across cosmic 
time \citep{Hamann1998,Warner2003,Nagao2006,Shin2013,Du2014}. Interestingly, quasi-periodic
ejections have been found in normal galaxies by {\it eROSITA} \citep{Arcodial2021}, implying that 
stellar-mass black holes (BHs) do reside around SMBHs in galactic centers. Compact objects form binaries 
in the very dense gaseous environment of SMBH disks, leading to $\gamma$-ray bursts and gravitational wave 
(GW) bursts from galactic nuclear regions \citep{Cheng1999}.  The detection by  Advanced LIGO/Virgo of 
GWs from the mergers of stellar binary BHs (BBHs; e.g., \citealt{Abbott2016a,Abbott2016b,Abbott2017}) 
has renewed theoretical interest in this 
problem \citep{Bartos2017, McKernan2019, McKernan2020, Yang2019, Grobner2020, Samsing2020,
Secunda2020, Tanaga2020, Yang2020, Li2021,Kaaz2021}. The GW190521 event has garnered special attention, 
not only because of the large masses of the two constituent BHs (85 and 66 $\sunm$; \citealt{Abbott2020}), 
but also because the event was potentially hosted by the quasar SDSS J1249+3449 \citep{Graham2020,Palmese2021}.  
AGNs and quasars could be natural factories of high stellar-mass BBHs efficiently formed in situ in 
SMBH disks.

Compact objects deeply enveloped by the extremely dense gas of the SMBH disk form a new kind of 
stellar population. Since their fates are modified by accretion from the massive envelope, we call them 
accretion-modified stars (AMSs). This general terminology covers a wide range of possible cores, ranging 
from main sequence stars \citep{Cantiello2020}, white dwarfs, neutron stars to BHs. The massive 
envelope of an AMS generally is associated with inflow from the Bondi accretion \citep[][]{Bondi1952}. 
It should be noted that AMSs are different from Thorne-\.Zytkow 
objects (\citealt{Thorne1975,Thorne1977}), not only 
in terms of their core, which consists of a neutron star, but also in terms of the physics of their 
massive envelope. As discussed in \cite{Wang2021} (see also Section~2.1), AMS BHs (hereafter 
AMS BHs) are fed by hyper-Eddington accretion\footnote{Usually when accretion rates exceed 
$(10^{3}\sim 10^{4})L_{\rm Edd}/c^{2}$, super-Eddington accretion is often called as hyper-Eddington 
accretion \citep[e.g.,][]{Takeo2020}.} with rates reaching up to $10^{9-10}\,L_{\rm Edd}/c^{2}$
for $10-10^{2}\sunm$ BHs, where $L_{\rm Edd}$ is the Eddington luminosity and $c$ is the speed of 
light. Such an accretion rate is much higher than the usual regime of slim accretion 
disks \citep{Abramowicz1988,Wang1999}. Hyper-Eddington accretion develops powerful 
outflows \citep[e.g.,][]{Takeo2020}, which have a profound effect on the evolution of AMSs. As 
described in \cite{Wang2021}, the outflows in AMSs are so strong that they can halt the accretion.  
The cumulative kinetic energy of the outflows drives an explosion. We call it Bondi explosion, 
which in a typical quasar manifests itself as a slow transient in the radio, optical-UV, soft X-ray, 
and $\gamma$-ray bands with an occurence rate of $\sim 1\,{\rm yr^{-1}}$ \citep[see Eq. 24 in][]{Wang2021}.  
On the other hand, AMSs trapped in a SMBH disk migrate radially with the accreting gas and gravitationally
interact with each other if they experience close encounters over several orbital periods. This results 
in the formation of AMS binaries, giving rise to additional phenomena such as GW bursts.

This paper explores the formation of AMS binaries as an unavoidable consequence in SMBH disks. The 
orbits of the binaries evolve through several phases until their final merger generates a GW burst.  
Newly born BHs from the mergers are rotating very fast due to orbital angular momentum (AM).
Therefore, three kinds of electromagnetic counterparts (EMCs) are considered due to: 1) Blandford-Znajek 
(BZ) power from the spin of the merged BHs \citep{Blandford1977}, 2) thermal emission from Bondi sphere 
and 3) non-thermal emission from Bondi explosion of AMS with BBHs. They have very different timescales 
for BHs of $100\sunm$. AGNs could be a factory of GW bursts. 

The paper is scheduled as follows. In Section 2, the formation of BBHs is investigated
based on the properties of the AMSs, and their formation rates are derived analytically. We study 
evolution of the binaries in \S3, in particular the observational appearance of EMCs of GW bursts 
when the binaries merge. Three kinds of EMCs driven by different mechanisms could appear as transients 
from radio to $\gamma$-rays. We draw conclusions in \S4.

\begin{figure*}
    \centering
    \includegraphics[width= 0.85\textwidth,trim=-15 170 10 45,clip]{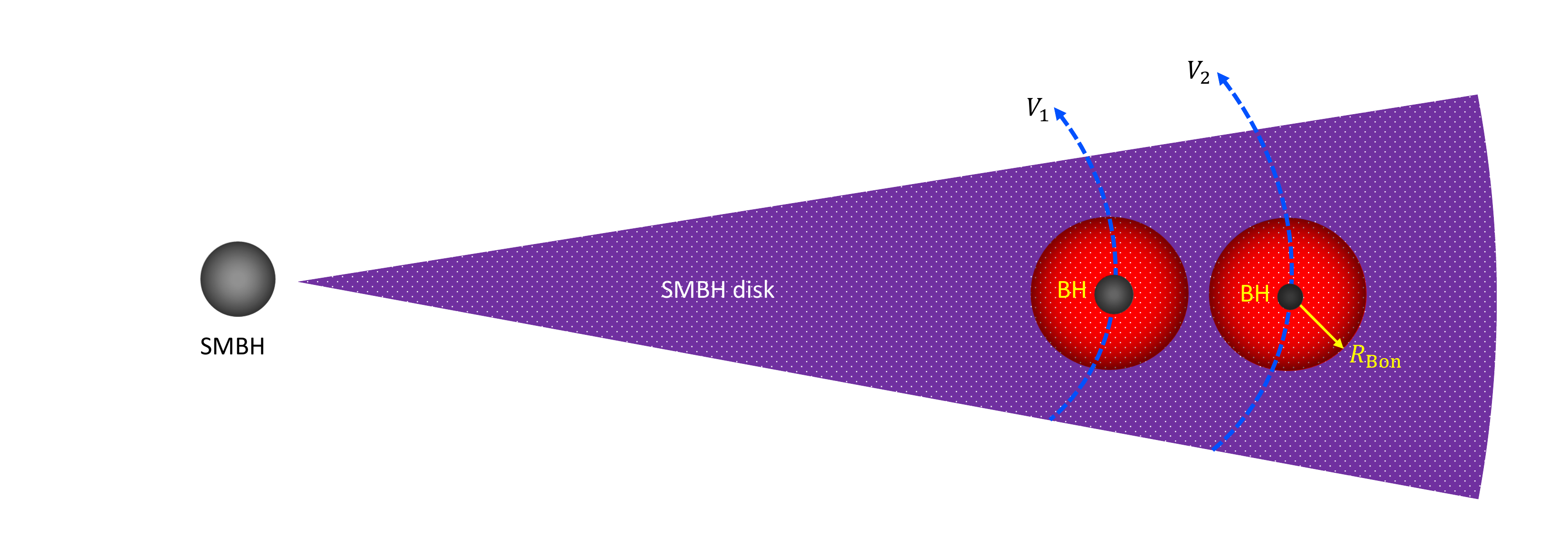}
    \caption{Top view of the SMBH disk. Surrounded by the SMBH disk cold gas, stellar-mass BHs form 
    AMSs through hyper-Eddington accretion. They are denoted type~I AMS. Bondi explosion of the AMS 
    creates cavities with a high-temperature and low-density medium (red), but the BHs remain there 
    and are still accreting with very low rates, forming type~II AMS. Pressure balance remains between 
    the cavity and the cold gas of the SMBH disk. The cavities are orbiting the central SMBH with 
    velocities $V_{1}$ and $V_{2}$, respectively. The differential velocity $|V_{2}-V_{1}|$ determines
    the timescale of binary formation once they encounter after many orbits around the central SMBH. 
    BBHs are formed in the cavities initially appearing as type~II AMS. Subsequent evolution of the 
    BBH orbit undergoes three different phases, as detailed in the text.}
\end{figure*}

\section{Formation of binary AMSs}
\subsection{Two types of AMS}
Compact objects will be formed through rapid evolution of massive stars in SMBH disks, which originate 
either from captures of stars from nuclear star clusters 
\citep{Artymowicz1993,Cheng1999,Cantiello2020},  or from star formation in the SG disks 
\citep{Collin1999,Collin2008,Goodman2003,Wang2010,Wang2011,Wang2012}. In this paper, we focus on
AMS BHs, whose properties depend on the mass density of the SMBH disks. Since the SG region of AGN 
accretion disks is still poorly understood, we continue to use the solution of the outer part of the 
standard accretion disk as the characteristic structure for discussions of AMSs and related issues. 
The half-thickness, density, mid-plane temperature, and radial velocity of the SMBH disk are 
\begin{equation}\label{Eq:disk}
\left\{
\begin{array}{l}\vspace{1ex}
H    =4.3\times 10^{14}\,\alpha_{0.1}^{-1/10}M_{8}^{9/10}\mathdotM^{3/20} r_{4}^{9/8}\,{\rm cm},\\ \vspace{1ex}
\rho_{\rm d} =6.9\times 10^{-11}\,\left(\alpha_{0.1}M_{8}\right)^{-7/10}\mathdotM^{11/20}r_{4}^{-15/8}\,{\rm g\,cm^{-3}},\\ \vspace{1ex}
T_{c}=4.6\times10^{3}\,\left(\alpha_{0.1}M_{8}\right)^{-1/5}\mathdotM^{3/10}r_{4}^{-3/4}\,{\rm K},\\
v_{r}=2.6\times 10^{2}\,\alpha_{0.1}^{4/5}M_{8}^{-1/5}\mathdotM^{3/10}r_{4}^{-1/4}\,{\rm cm\,s^{-1}},
\end{array}\right.
\end{equation}
respectively  \citep[e.g.,][]{Kato2008}.  Here the dimensionless quantities are the viscosity 
parameter $\alpha_{0.1}=\alpha/0.1$, the gravitational radius of the SMBH disk to that of the 
SMBH $r_{4}=R/10^{4}\Rg$, the gravitational radius $\Rg=G\BHm/c^{2}$, the gravitational constant 
$G$, and $M_{8}=\BHm/10^{8}\,\sunm$ is the central SMBH mass in units of $10^{8}\,\sunm$. The 
dimensionless accretion rate of the central SMBH is defined by 
$\mathdotM=\dot{M}_{\bullet}/\dot{M}_{\rm Edd}$, where $\dot{M}_{\rm Edd}=L_{\rm Edd}c^{-2}$ is 
the Eddington rate, with  the Eddington luminosity $L_{\rm Edd}=1.3\times10^{46}M_{8}\,{\rm erg\,s^{-1}}$, 
and $\dot{M}_{\bullet}$ is the accretion rate of the central SMBH. The Toomre (1964) parameter, 
defined as $Q=\Omega_{\rm K} c_{s}/\pi G \rho_{\rm d} H$, describes the disk self-gravity, where 
$c_{s}$ is the local sound speed of the disk ($c_{s}\approx 
15.7\,T_{4}^{1/2}\,\kms$), and $\Omega_{\rm K}=\sqrt{G\BHm/R^{3}}$.
The disk becomes SG beyond a critical radius where $Q=1$, which is given by
$R_{\rm SG}/\Rg=1.2\times 10^{3}\,\alpha_{0.1}^{28/45}M_{8}^{-52/45}\mathdotM^{-22/45}$.
We consider AMS in the region beyond $R_{\rm SG}$.
As shown by \cite{Wang2021}, most AMSs are trapped by and corotate with the gas in the SMBH disk.
The AMS BHs undergo episodic hyper-Eddington accretion driven by powerful outflows, leading to a 
Bondi explosion and maintaining very low-level accretion onto the BHs during every episode. 
AMS BHs with hyper-Eddington and low-accretion rates are denoted as type~I and type~II AMS.
The basic properties of type~I AMS can be estimated from the Bondi accretion of cold gas in the 
SMBH disk, as described in \cite{Wang2021} and below.

However, the BHs are still accreting from the hot, post-shock gas in the cavity of the SMBH disk 
after the Bondi explosion. The hot gas in the cavity determines the orbital evolution of the BBH. 
As shown in \citealt{Wang2021} (see Equation~\ref{Eq:Ekin}), the cavity radius
\begin{equation}\label{Eq:Rexp}
R_{\rm exp}=9.0\times 10^{15}\,E_{52}^{1/4}\alpha_{0.1}^{1/5}M_{8}^{9/20}
             \mathdotM^{-7/40}r_{4}^{15/16}\,{\rm cm},
\end{equation}
where $E_{52}=E_{\rm out}/10^{52}\,{\rm erg}$ is the kinetic energy of the outflow from the 
hyper-Eddington accretion of type~I AMS. The temperature of the shock-swept medium is
\begin{equation}
T_{\rm cav}\approx \frac{2(\Gamma_{\rm ad}-1)m_{\rm prot}}{(1+\Gamma_{\rm ad})^{2}k}V_{\rm exp}^{2}
 =2.3\times 10^{11}\,V_{\rm exp,5}^{2}\,{\textrm{K}},
\end{equation}
where $V_{\rm exp,5}=V_{\rm exp}/10^{5}\,{\rm{km\,s^{-1}}}$, $m_{\rm prot}$ is the mass of the
proton, $k$ is the Boltzmann constant, and we take the adiabatic index $\Gamma_{\rm ad}=5/3$. 
Since the vertical direction is open, most of the gas in the cavity will escape from the SMBH disk. 
In order to estimate the gas density of the cavity, we use the pressure balance between the cavity 
and the cold disk,
\begin{equation}
n_{\rm cav}=\frac{n_{\rm d}T_{\rm c}}{T_{\rm cav}}
           =4.4\times 10^{6}\,n_{\rm d, 14}T_{\rm d,4}V_{\rm exp,5}^{-2}\,{\rm cm^{-3}},
\end{equation}
where $n_{\rm d,14}=n_{\rm d}/10^{14}\,{\rm cm^{-3}}$ and $T_{\rm c,4}=T_{\rm c}/10^{4}\,$K are the 
number density and temperature of the SMBH disk, respectively, which allows us to estimate the 
accretion rate of the type~II AMS and their mass. It should be noted that the cooling timescale of 
$t_{\rm ff}\approx 10^{3}\,T_{11}^{1/2}n_{6}^{-1}\,$yr is comparable with the rejuvenation timescale 
of type~I AMS (see Eqn \ref{eq:trej}), where $T_{11}=T_{\rm cav}/10^{11}\,$K and 
$n_{6}=n_{\rm cav}/10^{6}\,{\rm cm^{-3}}$. The Bondi accretion rate can be expressed simply by
$\dotMBon=4\pi G^{2}\bhm^{2}\rho_{\rm d}/c_{s}^{3}$ for AMSs of both types, but with different
surrounding density and temperature, where $\bhm$ is the AMS BH mass (i.e., $\bhmp$ and $\bhms$). 
In this paper, we give characteristic values of the AMS for BHs with $10^{2}\sunm$ for a brief 
application to GW190521. Its dimensionless rates are
\begin{equation}\label{Eq:dotm}
\dot{m}_{\rm Bon}=\frac{\dotMBon}{\dot{M}_{\rm Edd}}\approx\left\{\begin{array}{ll}\vspace{1ex}
8.9\times 10^{9}\,m_{2}(\alpha_{0.1}M_{8})^{-2/5}\mathdotM^{1/10}r_{4}^{-3/4}&({\textrm{type~I AMS}}), \\
6.3\times 10^{-9}\,m_{2}n_{\rm cav,7}V_{\rm exp,5}^{-3} &({\textrm{type~II AMS}}),
\end{array}\right.
\end{equation}
where $m_{2}=\bhm/10^{2}\,\sunm$ and $n_{\rm cav,7}=n_{\rm cav}/10^{7}\,{\rm cm^{-3}}$. The Bondi 
radius of type I AMS given by $R_{\rm Bon}=G\bhm/c_{s}^{2}$, however, is limited by the tidal force 
of the SMBH. For $\bhm=10^{2}\sunm$, we find the Bondi radius
$R_{\rm Bon}=
1.2\times 10^{16}\,m_{2}\left(\alpha_{0.1}M_{8}\right)^{1/5}\mathdotM^{-3/10}r_{4}^{3/4}\,{\rm cm}$,
which is about 10 times the thickness of the SMBH disk, and thus have tidal-limited radius
\begin{equation}
\RBon= \left\{\begin{array}{ll}\vspace{1ex}
H &({\textrm{type~I AMS})},\\
R_{\rm exp} &({\textrm{type II AMS}}).
\end{array}\right.
\end{equation}
The Bondi sphere, limited by the tidal force of the SMBH, has the maximal height of the SMBH disk.
Here only the sound speed appears in the expression for the Bondi accretion rate; the relative 
velocity between the BH and SMBH disk is neglected because of corotation \citep[see Eq. 7 in][]{Wang2021}.
Moreover, separations of the binary AMSs are much smaller than the Bondi radius when the relative 
velocities of the two BHs are larger than the sound speed of the SMBH disks. Viscosity very efficiently
dissipates the orbital AM of the formed binary BHs (see Eq.\ref{Eq:tvis}). The validity of the Bondi 
accretion approximation is guaranteed by  $\RBon\ll R_{\rho}$, where 
$R_{\rho}=\left|d\ln\rho_{\rm d}/dR\right|^{-1}\approx R$ (i.e., $e$-folding variations of the disk 
density over $R_{\rho}$) is the density scale of the SMBH disk. Since the Bondi accretion of type~I 
AMS is hyper-Eddington, powerful outflows develop from the slim accretion 
disk \citep{Ohsuga2005,Kitaki2018}. Radiative feedback, which operates in super-Eddington 
accretion \citep{Wang2006,Milos2009a,Milos2009b}, may be dominated by outflows in hyper-Eddington 
accretion \citep{Takeo2020}. In the present context, we consider powerful outflows as the dominant 
mechanism that drives episodic accretion of the AMS BHs \citep{Wang2021}. 

The Bondi mass, which is defined as the gas mass within the Bondi radius, can be approximated by 
\begin{equation}\label{Eq:MBon}
M_{\rm Bon}=\frac{4\pi}{3} R_{\rm Bon}^{3}(\rho_{\rm d},\rho_{\rm cav}) =
\left\{\begin{array}{ll}\vspace{1ex}
2.1\times 10^{2}\sunm\,\rho_{\bar{10}}H_{15}^{3}&{\textrm{(type~I AMS)}},\\
3.5\times 10^{-6}\sunm\,R_{\rm Bon,15}^{3}n_{6}&{\textrm{(type~II AMS)},}
\end{array}\right.
\end{equation} 
where $\rho_{\bar{10}}=\rho_{\rm d}/10^{-10}\,{\rm g\,cm^{-3}}$, $H_{15}=H/10^{15}\,{\rm cm}$ and
$R_{\rm Bon,15}=R_{\rm Bon}/10^{15}\,{\rm cm}$.
The tidal force limits the size of the Bondi sphere and hence its mass similar to that of a $10\sunm$
BH \citep{Wang2021}. Without the tidal limit, the Bondi sphere of a $100\sunm$ BH
will be $10^{2}$ times that given by Eqn (\ref{Eq:MBon}).
A type~II AMS has much lower accretion rate and mass compared to its type~I counterpart. Type~II AMSs are 
expected to contain an advection-dominated accretion flow \citep{Narayan1994} and are generally too 
faint to be observed. 
On the other hand, the cavity formed by the Bondi explosion is replenished by the infall of gas from the 
SMBH disk, rejuvenating the AMS on a timescale of 
\begin{equation}\label{eq:trej}
t_{\rm rej}=\frac{R_{\rm exp}}{c_{s}}
   =267.0\,\alpha_{0.1}^{3/10}M_{8}^{11/20}\mathdotM^{-13/40}r_{4}^{21/16}E_{52}^{1/4}\,{\rm yr}.
\end{equation}
We then have the duty cycle of hyper-Eddington accretion episodes of the AMS BHs, 
$\delta_{\bullet}\approx 5.3\times 10^{-5}$, namely, the ratio of hyper-Eddington accretion time 
to the rejuvenation time \citep[see details in][]{Wang2021}.

Here it is helpful to distinguish between two kinds of hyper-accreting cases.
Hyper-Eddington accretion onto the BHs produces non-relativistic but powerful 
outflows, or mildly moving blobs from the choked jet if the BH is rotating maximally. 
This differs from the case of long $\gamma$-ray bursts \citep{Woosley1993}, whose 
highly relativistic jets are produced by accretion of neutrons (also some stellar envelope gas) 
onto BHs at hyper-Eddington rates. The cores of massive 
stars, where neutrino cooling dominates, typically supply an accretion rate of 
$1\,\sunm\,{\rm s^{-1}}\approx 10^{15}\,\dot{M}_{\rm Edd}$.  Unlike long $\gamma$-ray bursts, jet 
production could be suppressed in type~I AMS BHs, despite their hyper-Eddington accretion rates. 
This is evidenced by the fact that AGNs with high accretion rates are usually 
radio-quiet \citep[e.g.,][]{Ho2002,Ho2008,Sikora2007}, for 
jets are quenched in BHs accreting in their high, soft states \citep[e.g.,][]{Fender2004}.
However, the current situation for merging BHs with $\gtrsim 10^{9}\dot{M}_{\rm Edd}$
is uncertain based on the latest numerical simulations \citep{Sadowski2015,Sadowski2016}.
Usually neutrino cooling, which is extremely sensitive to temperature  [its rate is proportional 
to $(T/10^{11}\,{\rm K})^{9}$], is triggered when the temperature is higher than 
$10^{11}\,$K \citep[e.g.,][]{Popham1999}.  However, accretion rates of $10^{9-10}\,\dot{M}_{\rm Edd}$ 
are still not high enough to trigger neutrino cooling, since the overall temperature of a 
self-similar disk is only $\lesssim 10^{9}\,$K \citep{Wang1999}, and the temperature will be 
significantly lower if strong outflows are developed. Under such conditions we expect powerful 
outflows \citep{Takeo2020}, which have much wider opening angles than jets. 

On the other hand, strong magnetic fields play a key role in the formation of the highly 
relativistic jets in $\gamma$-ray bursts. In the same spirit, we explore the possibility that 
powerful relativistic jets could be produced either by the radiation pressure of super-Eddington 
accretion of non-rotating BHs \citep{Sadowski2015} or by the BZ mechanism of 
fast-rotating BHs \citep{Blandford1977}. Considering the many uncertainties of hyper-Eddington
accretion and the lack of clear observational tests, we explore both BZ-powered jets and
Bondi explosions as potential mechanisms for generating an EMC.

\subsection{Binary AMSs}
Three cases of binary AMS are possible: (1) type~I$+$I; (2) type~I$+$II, and (3) type~II$+$II. 
Considering that the duty cycle of type~I AMS is only $\sim 5\times 10^{-5}$, we expect
most AMSs to be type~II. While cases 1 and 2 are possible, their numbers are much smaller than 
those of case 3. In this paper, we only focus on the case where both members of the binary are 
type~II AMS.

Since AMSs are trapped by the SMBH disk, they migrate with the gas and form binaries through their 
gravitational interaction once they are sufficiently close. Given $N_{\bullet}$ BHs in the 
SMBH disk, their surface density is $\Sigma_{\bullet}=N_{\bullet}/\pi R^{2}$.  Considering velocity 
differences of $\Delta V=\Omega_{\rm K}\calA_{0}/2$, AMSs will encounter within a timescale of 
$t_{\rm bin}=2\pi R/n_{\bullet}\Delta V$ after many orbits around the central SMBH, where 
and $n_{\bullet}=2\pi R\calA_{0}\Sigma_{\bullet}$ is the number of BHs in an annulus of width 
$\calA_{0}$. In principle, an AMS binary can form when the gravity between the individual BHs is 
stronger than the tidal force of the SMBH acting on the  binary. Given the tidal force 
$F_{\rm tid}=G\BHm\bhm(\calA_{0}/R)/R^{2}$ and the gravity of the BBH 
$F_{\rm bin}=G\bhm^{2}/\calA_{0}^{2}$, the condition $F_{\rm bin}\ge F_{\rm tid}$ places an upper 
limit on the distance between two AMSs,
\begin{equation}\label{eq:a0}
a_{0}=\left(\frac{\bhm}{\BHm}\right)^{1/3}
                  \left(\frac{R}{r_{\rm g}}\right)
                  =1.0\times 10^{8}\,M_{8}^{2/3}m_{2}^{-2/3}r_{4},
\end{equation}
where $a_{0}=\calA_{0}/r_{\rm g}$ and $r_{\rm g}=G\bhmp/c^{2}$. This condition
can be also derived from the virial relation, that the sum of the kinetic and potential energy 
of the BBH vanishes. As shown by Equation~(\ref{eq:a0}), the upper limit of the separation between 
the two BHs is much smaller than the Bondi radius. This validates the approximation of 
Bondi accretion onto the BBH during the orbital evolution. Moreover, the relative velocities 
of the two BHs are
only $\sim 30\,{\rm km\,s^{-1}}$ initially, which is much smaller than the sound 
speed of the SMBH disk and cavity.  The bounded binary BHs remain inside a type II AMS
corotating with the gas of the SMBH disk\footnote{In the scenario of \cite{Cantiello2020}, 
stars in the SMBH disk are from captures from a nuclear star cluster. The BHs from these stars could
have different dynamics. This is beyond the scope of the present paper.}. The appearance of GW 
bursts and EMCs depends on the evolution of the BBH orbit and details of its accretion history.

For simplicity, we only discuss BHs of equal mass.  Once BHs enter the annulus $R-(R-\calA_{0})$ 
of the SMBH disk, BBHs form nearly instantaneously compared with the AGN lifetime, 
$t_{\rm AGN}\approx R/v_{r}\approx 1.8\times 10^{7}\,\alpha_{0.1}^{-4/5}M_{8}^{6/5}\mathdotM^{-3/10}r_{4}^{5/4}{\rm yr}$ (from Eq. 1).
For binaries formed from neighbouring BHs, the timescale for BBH formation is
\begin{equation}
t_{\rm bin}=\frac{4\pi}{n_{\bullet}\Omega_{\rm K}}\left(\frac{\BHm}{\bhm}\right)^{1/3}
           =2.5\times 10^{4}\,N_{40}^{-1}M_{8}^{5/3}m_{2}^{-2/3}r_{4}^{3/2}\,{\rm yr},
\end{equation}
in the formation zone, with a rate from one quasar
\begin{equation}\label{eq:Nbin}
\dot{N}_{\rm bin}=\frac{n_{\bullet}}{t_{\rm bin}}\left(\frac{R}{\calA_{0}}\right)
                 =3.2\times 10^{-3}\,N_{40}^{2}m_{2}^{2/3}M_{8}^{-5/3}r_{4}^{-3/2}\,{\rm yr^{-1}},
\end{equation}
where $N_{40}=N_{\bullet}/40$ from the entire SMBH disk\footnote{Wang et al. (2021) estimate the 
number of $10\,\sunm$ BHs, but the number of $10^2\sunm$ BHs is hard 
to estimate for a number of reasons, including uncertainties in the initial mass function of the 
progenitor stars and the growth of the BHs in the SMBH disk. Considering the possibility 
that the initial mass function in the disk might be top-heavy, we
assume $N_{\bullet}\propto \bhm^{-1}$ and have $N_{\bullet}=40$ BHs with $10^{2}\sunm$.}, 
and the factor $R/\calA_{0}$ is the number 
of formation zones. The rate in Equation (11) is an order of magnitude higher than the rate for tidal 
disruption events of stars in normal galactic centers \citep[e.g.,][]{Rees1988}.  The estimate of 
$N_{\bullet}$ depends on the number of massive stars in the SMBH disk, whose formation  efficiency 
relies on the initial mass function. Here we conservatively estimate $N_{\bullet}$ assuming a star 
formation efficiency of $0.1$ and a top-heavy initial mass function with a power-law index of $0.5$ 
(see \citealt{Wang2021}). In this paper, we neglect AMSs composed of neutron stars, which likely undergo 
more complicated processes than BHs. 

We note that number of BHs in the SMBH disk could be decreased by BBH mergers. For the binary 
rates given by Equation~(\ref{eq:Nbin}), $dN_{\bullet}/dt=-\dot{N}_{\rm bin}$ implies 
$N_{\bullet}=N_{0}/\left[1+\pi^{-1}\Omega_{\rm K}(\bhm/\BHm)^{2/3}N_{\bullet}^{0}t\right]$, 
where $N_{0}$ is the initial number of BHs. The asymptotic evolution of the number 
of BHs is $N_{\bullet}\propto t^{-1}$ when 
$t\gg N_{0}^{-1}\pi \Omega_{\rm K}^{-1}(\bhm/\BHm)^{-2/3}$.  Here we neglect the growth of the 
merging BHs. This binary rate is the maximum value for the initially given number of BHs.

\begin{sidewaysfigure*}\label{fig:orbit}
    \centering
\includegraphics[width= 0.95\textwidth,trim=40 45 60 10,clip]{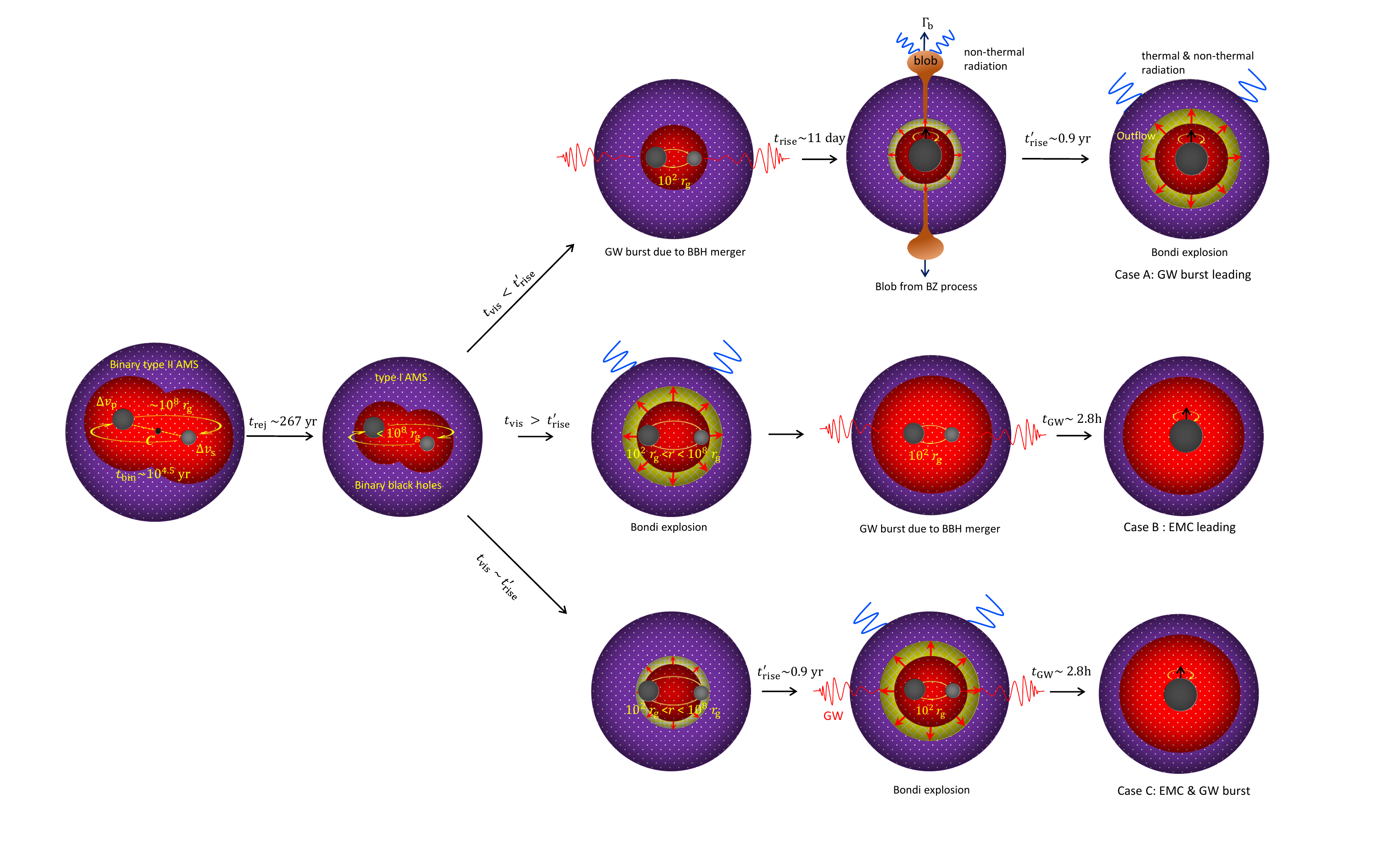}
\caption{A binary black hole (BBH), with primary mass $m_{\rm p}$, secondary mass $m_{\rm s}$, and 
center of mass $C$, is formed while cold gas from the SMBH disk (purple) replenishes the cavity (red). 
All numbers of this cartoon are for $100\,\sunm$ BHs.
A type~I AMS rejuvenates from a type~II AMS with a timescale $t_{\rm rej}$. The orbital evolution 
of the BBH is initially controlled 
by the tidal torque of the central SMBH, later by hyper-Eddington accretion, and finally the merger 
of the two BHs because of GW radiation. The spin AM of the Bondi sphere is very low 
because of efficient spin-down by tidal torque of the central SMBH. Three kinds of EMCs can 
be generated: 1) ejected blob moving with mildly relativistic velocity (its Lorentz factor 
$\Gamma_{\rm b}\sim 2$) by the Blandford-Znajek mechanism pumping the spin energy of the merged 
BHs; 2) Thermal emission from Bondi sphere (typically 4200\,K); and 3) Non-thermal emission from 
Bondi explosion driven by the outflows developed from the hyper-Eddington accretion.
Compared with the blob, Bondi explosion gives rise to a slow transient as an EMC peaking in 
optical bands, but non-thermal emission from the shocked gas can also arise in the radio band to 
energy of a few TeV. Here we stress that the three possible cases depend on
the degree to which viscosity removes the orbital AM of the BBHs.}
\end{sidewaysfigure*}

\section{Mergers of binary black holes}
\subsection{Orbital evolution of BBHs}
Differential rotation of the SMBH disk gives rise to initial orbital AM for the 
binary type~II AMS co-rotating with the disk. As shown in Figure~2, 
the velocities of the primary and secondary BHs relative to the center of mass are
$\Delta v_{\rm p}=V_{\rm K}a_{\rm p}/2R$ and $\Delta v_{\rm s}=-V_{\rm K}a_{\rm s}/2R$, respectively, 
where $V_{\rm K}=R\Omega_{\rm K}$ is the Keplerian velocity,
$a_{\rm p}=q\calA_{0}/(1+q)$ and $a_{\rm s}=\calA_{0}/(1+q)$ are the distances of the primary 
and secondary BHs to the center, $\calA_{0}$ is the initial separation of the BBH, and 
$q=\bhms/\bhmp$ is the mass ratio of the binary. Here the negative sign of $\Delta v_{\rm s}$ 
means that the two BHs have opposite velocities in the center-of-mass frame. The initial orbital 
AM is then given by  
$\calJ_{0}=\bhmp V_{\rm K}\left(\calA_{0}^{2}/2R\right)q(1+q)^{-1}$.
For a BBH in Keplerian rotation, its circular orbital AM is 
$\calJ_{\rm cir}=(G\calA)^{1/2}\bhmp^{3/2} q/(1+q)^{1/2}$, and 
$\calA=(\BHm/\bhmp)(\calA_{0}/R)^{3}\calA_{0}/4(1+q)\sim \calA_{0}$ from 
$\calJ_{0}= \calJ_{\rm cir}$ for $\BHm=10^{8}\sunm$, $\bhmp=10\sunm$, $\calA_{0}=10^{15}\,$cm, and 
$R=10^{4}\,R_{\rm g}$ from $\calJ_{\rm cir}=\calJ_{0}$. The initial BBHs are expected to have circular 
orbits. Figure 2 outlines the evolutionary track of the BBH from its birth.

The SMBH exerts a tidal torque on the BBHs given by 
\begin{equation}\label{eq:Ttid}
\calT\approx V_{\rm K}^{2}\bhmp(1+q)\left(\frac{\calA}{R}\right)^{2}
\end{equation}
on a tidal timescale of
\begin{equation}
t_{\rm tid}=\frac{\calJ_{\rm cir}}{\calT}
           =15.7\, q(1+q)^{-3/2}a_{8}^{-3/2}m_{2}^{-1}M_{8}^{2}r_4^{3}\,{\rm yr},
\end{equation}
where $a_{8}=a/10^{8}$ and $a=\calA/r_{\rm g}$. During its orbit evolution, a type~II AMS undergoes 
rejuvenation and accumulates gas through Bondi accretion. Compared with the rejuvenation timescale of 
AMS, the tidal timescale will be longer than $t_{\rm rej}$ after $a\lesssim 10^{7}$.   
In other words, the tidal torque does not efficiently remove orbital AM of the BBHs after a 
period of $t_{\rm tid}\sim t_{\rm rej}$. Since the subsequent process is much more efficient 
than tides, we neglect tidal effects such as the evolution of eccentricity and semi-major axis by the 
Kozai-Ledov mechanism \citep[e.g.,][]{Naoz2016}.  However, accretion onto the BBH should be considered 
\citep[e.g.,][]{Antoni2019,Comerford2019}.  We approximate the accretion as if the binary were
a single BH because the binary separation is much smaller than the Bondi radius. In the future, 
the orbital evolution of the BBH should consider accretion onto each BH.

An AMS with a single BH should have very low spin AM. The spin AM of the Bondi spehere
from the differential rotation of the SMBH disk is given by 
\begin{equation}
\calJ_{\rm Bon}\approx \MBon V_{\rm K}\frac{\RBon^{2}}{2R}.
\end{equation}
However, an AMS trapped in the SMBH disk will be synchronized with the orbital rotation by the tidal 
torque.  This is just opposite to the AM of the accreted gas from the disk. Similar to 
Equation~(\ref{eq:Ttid}), the tidal torque exerting on the Bondi sphere, 
$\calT^{\rm Bon}\approx V_{\rm K}^{2}M_{\rm Bon}(R_{\rm Bon}/R)^{2}$, removes spin AM of the Bondi 
sphere on a timescale 
\begin{equation}
t_{\rm tid}^{\rm Bon}=\frac{\calJ_{\rm Bon}}{\calT^{\rm Bon}}=0.5\Omega_{\rm K}^{-1}
                     =7.8\,r_{4}^{3/2}M_{8}\,{\rm yr}.
\end{equation} 
This indicates that the tidal torque efficiently removes the spin AM. We thus expect that 
the Bondi sphere has very low spin AM. After the tidal interaction phase with the SMBH,  
the BBH enters the rejuvenation phase during which it will undergo accretion from the Bondi sphere 
and efficiently remove the BBH orbital AM. 

Assuming that the viscosity in the Bondi sphere follows the standard $\alpha$ prescription 
\citep{Shakura1973}, the torque due to viscosity is given by 
${\cal T}_{\rm vis}=\alpha \rho_{\rm Bon} c_{s}^{2}{\mathscr V}_{\rm Bon}=\alpha E_{\rm th}$, where 
${\mathscr V}_{\rm Bon}$ is the volume of the Bondi sphere and 
$E_{\rm th}=\rho_{\rm Bon}c_{s}^{2}{\mathscr V}_{\rm Bon}$ is the thermal energy of the sphere. Since 
the BBH is accreting from the sphere at a hyper-Eddington rate, 
powerful outflows heat the Bondi sphere with an energy
\begin{equation}\label{Eq:Ekin}
E_{\rm out}=L_{\rm out}t_{\rm a}=1.3\times 10^{52}\,\eta_{0.1}f_{\bar{3}}m_{2}\dot{m}_{10}t_{\rm a,6}\,{\textrm{erg}},
\end{equation}
where $L_{\rm out}=\eta f_{\rm a}\dot{m}_{\rm Bon}L_{\rm Edd}$ is the kinetic energy of the outflows, 
$f_{\rm a}=10^{-3}f_{\bar{3}}$ is the fraction of the Bondi flow to fall into the BH, 
$\dot{m}_{10}=\dot{m}_{\rm Bon}/10^{10}$,
$\eta_{0.1}=\eta/0.1$ is the efficiency of accretion, and $t_{\rm a,6}=t_{\rm a}/10^{6}\,{\rm s}$ 
is the episodic accretion timescale (depending on some other parameters of the SMBH disk). We 
conservatively assume that the thermal energy of the
Bondi sphere is $E_{\rm th}=\alpha E_{\rm out}=10^{51}\,\alpha_{0.1}E_{52}\,{\rm erg}$ from
the kinetic outflows. This can be justified by the role of turbulence in the thermalization of 
the kinetic energy of the outflows. The timescale for removing orbital AM of the BBH,
\begin{equation}\label{Eq:tvis}
t_{\rm vis}=\frac{\calJ_{\rm cir}}{{\cal T}_{\rm vis}}
  =0.28\,\alpha_{0.1}^{-2}E_{52}^{-1}m_{2}^{2}a_{8}^{1/2}q(1+q)^{-1/2}\,{\rm yr}.
\end{equation}
It should be noted that this timescale is very sensitive to $\alpha$, which is quite uncertain. 

After rejuvenation, the accretion rates of type I AMSs reach $\sim 10^{9-10}L_{\rm Edd}c^{-2}$ (see 
Eqn \ref{Eq:dotm}), and powerful outflows will be developed \citep{Ohsuga2005,Kitaki2018,Takeo2020}. 
The outflows have strong impact on the hyper-Eddington accretion through strong shocks heating the 
AMS. The accretion terminates if the temperature of the post-shocked medium is higher than the 
virial temperature, and the hyper-Eddington accretion is thus episodic with a timescale ($t_{\rm a}$). 
See details in \cite{Wang2021}. The accretion episode of the AMS, approximated by the model for a 
single BH, occurs on a timescale
\begin{equation}\label{Eq:ta}
t_{\rm a}=\left(\frac{16\sqrt[3]{5}}{9}
          \frac{c^{4/3}r_{\rm g}^{2/3}M_{\rm Bon}}{\alpha^{2/3}L_{\rm out}}\right)^{3/5}
           =4.5\times 10^{5}\,\eta_{0.1}^{-3/5}\alpha_{0.1}^{-2/5}
           f_{\bar{3}}^{-3/5}\dot{m}_{10}^{-3/5}m_{2}^{-1/5}(M_{\rm Bon,2}/2.1)^{3/5}\,{\textrm{s}},
\end{equation}
where $M_{\rm Bon,2}=M_{\rm Bon}/10^{2}\sunm$.

Subsequent evolution of the AMS with BBHs depends on the three timescales of $t_{\rm vis}$,
$t_{\rm GW}$ and $t_{\rm rise}$ ($t_{\rm rise}^{\prime}$), which are given by Eqn.(\ref{Eq:tvis},
\ref{Eq:tGW} and \ref{Eq:trise} or \ref{Eq:trisep}) in Section 3.3. Observational appearance of the 
EMCs is due to two kinds of mechanisms driven by the BZ process (depending on BH spin) and dynamics 
of the Bondi sphere (Eq. \ref{Eq:tGW}), respectively. Figure 2 shows a cartoon depicting three 
channels producing EMCs and GW bursts. Since the orbital evolution strongly depends on viscosity 
($\alpha$ in Eqn.\ref{Eq:tvis}), generally they are divided by the viscosity timescale of the Bondi 
sphere compared with the rise timescale of radiation from the sphere. Case A appears when the viscosity 
efficiently removes orbital AM, case B is inefficient, and case C is moderately efficient.

\subsection{BBH Mergers}
The initial separation of a BBH, $\calA_{0}=10^{8}r_{\rm g}$, will be greatly reduced by 
an amount that depends on the AM of the rest of the hot gas within the cavity after the Bondi 
explosion. Eccentricity evolution is also important for GW bursts \citep{Grobner2020,Secunda2020}, 
but we only take into account circular orbits of BBHs. 
GW process dominates when $t_{\rm vis}$ (given by Eq.\ref{Eq:tvis}) is longer than
the timescale for the BBH to merge due to GW radiation,  
\begin{equation}\label{Eq:tGW}
t_{\rm GW}=\frac{5a^{4}}{64q(1+q)}\left(\frac{r_{\rm g}}{c}\right)
         \approx 2.8\,(a_{2}/1.27)^{4}m_{2}q^{-1}(1+q)^{-1}\,{\rm hour},
\end{equation}
where $a_{2}=a/10^{2}$ \citep{Peters1964}, and the corresponding separation of the BBHs
\begin{equation}
a=126.9\,\alpha_{0.1}^{-4/7}q^{4/7}(1+q)^{1/7}E_{\rm 52}^{-2/7}m_{2}^{2/7}.
\end{equation}
The GW frequencies are 
\begin{equation}\label{Eq:fGW}
f_{\rm GW}=56.7\,(1+q)^{1/2}m_{2}^{-1}(a/5)^{-3/2}\,{\textrm{Hz}},
\end{equation}
which fall within the regime of LIGO. GW bursts due to the present mechanism occur at a detectable 
rate of
\begin{equation}
\dot{{\cal R}}_{\rm GW}\approx n_{\rm q}{\mathscr V}_{\rm cm}\dot{N}_{\rm bin}
                       \approx 3.0\, n_{\rm q,6}{\mathscr V}_{158}\dot{N}_{\bar{3}}\,{\rm yr^{-1}}, 
\end{equation}
where $n_{\rm q,6}=n_{\rm q}/6\,{\rm Gpc^{-3}}$ is the number density of quasars (for $z\approx 1$, see
\citealt{Richards2006}), ${\mathscr V}_{158}={\mathscr V}_{\rm cm}/158\,{\rm Gpc}^{3}$ is the 
co-moving volume within $z=1$, and $\dot{N}_{\bar{3}}=\dot{N}_{\rm bin}/3.2\times 10^{-3}\,{\rm yr}^{-1}$
(in one quasar). Here we assume cosmological parameters $\Omega_{m}=0.3$, $\Omega_{\Lambda}=0.7$, and
$H_{0}=70\,\kms\,{\rm Mpc^{-1}}$. 

The LIGO-Virgo collaboration estimates a BBH merger rate of $53.2_{-28.8}^{+58.5}\,{\rm Gpc^{-3}\,yr^{-1}}$ 
for the local Universe from the O1 and O2 observing runs \citep{LIGO2019a,LIGO2019b}. This rate can be 
explained by BBH mergers from dense star clusters \citep[e.g.,][]{Antonini2012,Martinez2020}. Our present 
BBH merger rate is lower than the detected rate by one order of magnitude, although the detection of 
GW190521 is consistent with the predictions of this paper. The present prediction is also lower than that of 
$12\,{\rm Gpc^{-3}\,yr^{-1}}$ by Secunda et al. (2020; see their Equation~7) and \cite{McKernan2019}, 
but they use a BH number of $N_{\bullet}=2\times 10^{4}$, much  larger than ours.  Moreover, 
the difference comes from the fact that the present scenario favors mergers of high-mass BBH in the 
very dense environment of a SMBH disk. Mergers of BBHs less than $10\,\sunm$ would be fainter, and 
their EMCs are hard to detect.

The abundance of AMS related to GW bursts in SMBH disks can be plausibly tested by AGN variability.
As predicted by \cite{Wang2021}, the characteristic light curves decay as $t^{-6/5}$, with no
intra-band delays from radio to $\gamma$-rays. Identifying these features of AGN light curves will 
greatly advance our understanding of AMS physics in the SMBH disk, as well as the role they play in 
supplying gas to the central engine.

\subsection{Electromagnetic counterpart}
{\cblue In this sub-section, 
we outline the characteristics of AMS emissions as EMCs of 
GW bursts. Emissions from the AMS depend on the details of the merged BHs and the Bondi sphere,
but also on the specifics of the broad-line region (BLR) of AGNs. Except for the Bondi explosion,
some energy pumped from the BH spin is released, as the AMSs must rotate very quickly from the 
orbital AM, whatever their spins {\textit{prior}} to the merger \citep[e.g.,][]{Hughes2003}. In 
this paper, we explore two possible channels for generating EMCs: 1) relativistic jets from the 
hyper-Eddington accretion driven by BZ mechanism pumping rotating energy of the BHs; 
and 2) Bondi explosion driven by outflows from the hyper-Eddington accretion. 

\subsubsection{Relativistic ejecta after BHs merge}
In Appendix, we derive the BZ power during the accretion episode of the BH in 
an AMS. The BZ-powered jet is choked by the dense medium of the Bondi sphere, forming a blob as ejecta 
from the Bondi sphere. In this paper, we neglect details of the choking processes in order to   
estimate the average Lorentz factor of the ejected blob in light of the energy equation given by 
$E_{\rm BZ}\approx E_{\rm kin}=2\Gamma_{\rm b} \Delta M_{\rm BZ}c^{2}$, which yields
\begin{equation}\label{Eq:Gamma}
\Gamma_{\rm b}=1.8\,E_{\rm BZ,51}R_{\rm jet,12}^{-2}R_{\rm Bon,15}^{-1}\rho_{\bar{10}},
\end{equation}
where $E_{\rm BZ,51}=E_{\rm BZ}/10^{51}\,{\rm erg}$ is the energy pumped from the BH spin,
$E_{\rm kin}$ is the kinetic energy of the blob, $R_{\rm jet,12}=R_{\rm jet}/10^{12}\,{\rm cm}$ 
is the jet radius, and  
$\Delta M_{\rm BZ}\approx \pi R_{\rm jet}^{2}R_{\rm Bon}\rho_{\rm d}
                  \approx 1.6\times 10^{-4}\,R_{\rm jet,12}^{2}R_{\rm Bon,15}\rho_{\bar{10}}\,\sunm$
is the mass contained in the volume of the jet configuration. 
Here the factor 2 accounts for the double-sided nature of
relativistic jets. Eqn.(\ref{Eq:Gamma}) shows that the ejected blobs are moving mildly 
relativistically, but Doppler boosting significantly enhances the luminosity by a factor 
of $\Gamma_{\rm b}^{4}$. The initial radius of the blob is approximated by 
$R_{\rm blob}^{0}\approx \left(R_{\rm jet}^{2}R_{\rm Bon}\right)^{1/3}
                 \approx 10^{13}\,R_{\rm jet,12}^{2/3}R_{\rm Bon,15}^{1/3}\,$cm, corresponding to 
the initial optical depth of  
$\tau_{\rm es}^{0}=\kappa_{\rm es}\rho_{\rm d}R_{\rm blob}^{0}
                  =340\,\rho_{\bar{10}}\left(R_{\rm blob}^{0}/10^{13}\,{\rm cm}\right)^{-2}$, 
where $\kappa_{\rm es}=0.34$ is the opacity of electron scattering. The total energy of the blob 
is composed of two components, $E_{\rm BZ}=E_{\rm kin}+E_{\rm th}$, where $E_{\rm th}$ is its thermal 
energy. The relative strength of the two components depends on the details of the 
dynamics of the choked jet inside the Bondi sphere. Meanwhile, collimation supported by the 
cocoon of the dense surrounding keeps the choked jet in the inner part of the SMBH 
disk \citep{Bromberg2011,Perna2021,Zhu2021}. Considering the opening angle 
$\theta\approx c_{\rm s}/c\approx \Gamma_{\rm b}^{-1}$, where $c_{s}$ is the sound speed of the choked jet,
we have $E_{\rm kin}\approx \Gamma_{\rm b}^{2}E_{\rm th}$ when the blob is ejected outside the SMBH 
disk (or atmosphere of the disk). The ratio $E_{\rm kin}/E_{\rm th}$ decreases with the expansion 
and motion of the blob, as it sweeps through the BLR after it is born.

The blob undergoes three phases: 1) free expansion; 2) blastwave, when the swept medium is 
comparable with the mass of the blob; and 3) snowplow, when the swept mass is about 
$\zeta\approx 10-30$ times the initial mass of the blob 
\citep[e.g., details in latest numerical simulations of][]{Petruk2021}. The characteristic
expansion velocity of the blob is approximated by
\begin{equation}
\frac{V_{\rm exp}}{c}=\left\{\begin{array}{ll}
1      &({\rm for}\,\, t\lesssim t_{c}),\\
       &     \\
1.0\,(t/t_{c})^{-3/5} &({\rm for}\,\, t\gtrsim t_{c}). \\
\end{array}\right. 
\end{equation}
Namely, the blob expands with the speed of light before $t_{c}$, and then with the
Sedov velocity, where $t_{c}\approx 2.9\times 10^{4}\,$s is given by $V_{\rm exp}/c=1$ and
$R_{\rm c}=ct_{\rm c}=8.8\times 10^{14}\,$cm. We find that 
$R_{\rm c}\approx R_{\rm exp}=(E/\rho)^{1/5}t_{c}^{2/5}$ (Sedov expansion) for $E=10^{51}\,$erg
and $\rho\approx 10^{-17}\,{\rm g\,cm^{-3}}$. At this moment, the ejected blob 
has an optical depth 
$\tau_{\rm es}=\rho_{\rm blob}\kappa_{\rm es}R_{\rm blob}
             \approx 10^{-2}\,\Delta M_{\rm BZ,\bar{4}}R_{\rm blob,15}^{-2}$,
where $\rho_{\rm blob}=3\Delta M_{\rm BZ}/4\pi R_{\rm blob}^{3}$ is the blob density,
$\Delta M_{\rm BZ,\bar{4}}=\Delta M_{\rm BZ}/10^{-4}\,\sunm$ and
$R_{\rm blob,15}=R_{\rm blob}/10^{15}\,{\rm cm}$.
The swept mass for $t\gtrsim t_{c}$ is given by
\begin{equation}
\frac{\Delta M_{\rm BLR}}{\sunm}\approx 2.4\times 10^{-5}\left(\frac{t}{t_{c}}\right)^{6/5}.
\end{equation}
As shown by numerical simulations, the kinetic energy can be converted into thermal energy when
the swept mass is about $\zeta=10-30$ times the mass of the blob \citep[e.g.,][]{Petruk2021}.
The luminosity peaks because no significant energy is supplied by the conversion of the 
kinetic energy to the blob, as thermal energy drives the expansion. For
$\Delta M_{\rm BLR}=15\,\zeta_{15} \Delta M_{\rm BZ}$, we have a rise timescale of
\begin{equation}\label{Eq:trise}
t_{\rm rise}\approx 10.7\,\zeta_{15}^{5/6}\Delta M_{\rm BZ,\bar{4}}^{5/6}
n_{\rm BLR,7}^{-1/3}E_{51}^{-1/2}\,{\rm day},
\end{equation}
where $\zeta_{15}=\zeta/15$, and $n_{\rm BLR,7}=n_{\rm BLR}/10^{7}\,{\rm cm^{-3}}$ is the number
density of the BLR. The observed luminosity of the blob
\begin{equation}
L_{\rm peak}^{\rm BZ}\approx \Gamma_{\rm b}^{4}\xi L_{\rm BZ}
           =3.2\times 10^{45}\,\Gamma_{2}^{4}\xi_{0.1}f_{\bar{3}}\dot{m}_{10}m_{2}\,{\rm erg\,s^{-1}},
\end{equation}
where $\Gamma_{\rm b}=2\Gamma_{2}$, and 
$\xi=0.1\xi_{0.1}$ is the radiative efficiency of the kinetic energy of the blob. This is 
consistent with the estimation from $\Gamma_{\rm b}^{4}\xi E_{\rm BZ}/t_{\rm rise}$, where $E_{\rm BZ}$
is the total energy pumped from the BH spin (see Appendix).
The rise luminosity is proportional to 
$d\Delta M_{\rm BLR}/dt\propto R_{\rm blob}^{2}n_{\rm BLR}\propto t^{4/5}$ (swept by the transverse 
motion of the blob).
After $t_{\rm rise}$, the blob is significantly slowed down, and no significant fraction 
of the kinetic energy of the blob is converted into thermal energy for expansion,
causing the radiation to decay. Considering that the luminosity radiated from the blob is from
the expansion, we have $L_{\rm obs}\propto V_{\rm exp}^{2}\propto t^{-6/5}$. 
We approximate the light curve as
\begin{equation}
L_{\rm obs}=\left\{\begin{array}{ll}
L_{\rm peak}^{\rm BZ}\left(t/t_{\rm rise}\right)^{4/5} & ({\textrm{for}}\,\, t\lesssim t_{\rm rise}),\\ 
     &   \\
L_{\rm peak}^{\rm BZ}\left(t/t_{\rm rise}\right)^{-6/5} & (\textrm{for}\,\, t\gtrsim t_{\rm rise}).\\
\end{array}\right.
\end{equation}
After one GW burst, the AMS first forms an ejected blob through the BZ mechanism (mainly determined 
by $t_{\rm a}$), and then the ejected blob appears as a giant flare with rise and decay timescales
of a few $\times 10^{6}\,$s. The EMC appears more than 10 days after the GW burst.

For a simple estimation of the spectral energy distribution, we follow the treatment in \cite{Wang2021}.
Relativistic electrons are produced by Fermi acceleration \citep{Blandford1987} but lose their energy 
via synchrotron and inverse Compton (IC) scattering. The maximum Lorentz factor is given by the balance 
between the acceleration and radiation. Shocks from the Sedov expansion accelerate electrons to the
relativistic regime, generating a flare of non-thermal radiation from radio to $\gamma$-ray energies. 
For a magnetic field in equipartition with the post-shocked gas, we have 
$B=6.0\left(n_{\rm BLR,7}T_{9}\right)^{1/2}\,$G, where $T_{9}=T/10^{9}\,$K is the temperature of the 
post-shocked gas. The maximum 
Lorentz factor of the electrons is determined by the balance between energy loss and gain. 
For a typical quasar with $\BHm=10^{8}\sunm$ and $\mathdotM=1.0$, the energy density of the radiation 
field is about $u_{\rm ph}=L_{\rm bol}/4\pi R_{\rm BLR}^{2}c=0.16\,L_{45}R_{50}^{-2}\,{\rm erg\,cm^{-3}}$,
which peaks at UV frequencies of $\nu_{\rm UV}\sim 10^{15}$Hz,
where $L_{45}=L_{\rm bol}/10^{45}{\rm erg\,s^{-1}}$ and $R_{50}$ is the BLR radius in units of 50 
light-days \citep{Wang2021}. We find $u_{\rm ph}\ll u_{\rm B}$, where $u_{\rm B}=B^{2}/8\pi$ is the
energy density of magnetic fields; namely, synchrotron radiation dominates
over IC. Taking the balance between the acceleration and synchrotron loss,
we have the maximum Lorentz factor of 
$\gamma_{\rm max}=1.5\times 10^{6}\,(B/6\,{\rm G})^{1/2}T_{9}^{1/2}$ at the peak time of
luminosity. The synchrotron and IC luminosities are
\begin{equation}
L_{\rm syn}^{\rm blob}=1.6\times 10^{45}\,\xi_{0.1}\Gamma_{2}^{4}E_{51}t_{\rm rise,6}^{-1}\,{\rm erg\,s^{-1}};\quad
L_{\rm IC}^{\rm blob}=1.6\times 10^{44}\,\Gamma_{2}^{4}\left(\frac{u_{\rm ph}/u_{\rm B}}{0.1}\right)
                                       \left(\frac{L_{\rm syn}^{\rm blob}}{10^{44}\,{\rm erg\,s^{-1}}}\right)\,
                                       {\rm erg\,s^{-1}},
\end{equation}
with the synchrotron and IC frequencies, respectively, where $t_{\rm rise,6}=t_{\rm rise}/10^{6}\,$s,
\begin{equation}\label{Eq:nu}
\nu_{\rm syn}=206.4\,\left(\frac{B}{6\,{\rm G}}\right)\Gamma_{2}\gamma_{\rm max,6}^{2}\,{\rm keV},\quad
\nu_{\rm IC}=8.2\,\Gamma_{2}\gamma_{\rm max,6}^{2}\nu_{\rm UV,15}\,{\rm TeV},
\end{equation}
where $\gamma_{\rm max,6}=\gamma_{\rm max}/10^{6}$ and $\nu_{\rm UV,15}=\nu_{\rm UV}/10^{15}\,$Hz.
After the luminosity peak, the characteristic frequencies shift toward lower frequencies with time. 
The ejected blob will release energy over a wide range of frequencies from radio to TeV. It should
be pointed out that all calculations for radiations keep at a level of estimations rather than an
self-consistent way. 

It should be noted that the BZ-powered EMC depends on a sufficiently high BH spin, and accretion 
rate, which implies that BZ mechanism does not work in type II AMSs. There is absence of BZ-powered
blobs in case B and C.
}

\subsubsection{Bondi explosion}
The huge energy 
accumulated from the hyper-Eddington accretion (Eqn\,\ref{Eq:Ekin}) during $t_{\rm a}$ will drive the Bondi 
explosion, which can be divided into three phases. As briefly discussed in \cite{Wang2021}, internal 
shocks due to the collision between the outflows and the Bondi sphere efficiently dissipate kinetic 
energy into thermal energy. 
The Bondi sphere freely expands, and then enters the blastwave and snowplow phases.
We approximate its expansion velocity
\begin{equation}
V_{\rm exp}=\left\{\begin{array}{ll}
V_{\rm free}\approx 3.2\times 10^{3}\,E_{52}^{1/2}M_{\rm Bon,2}^{-1/2}\,{\rm km\,s^{-1}}&({\rm for}\,\, t\lesssim t_{c}^{\prime}),\\
       &     \\
V_{\rm Sedov}=3.2\times 10^{3}\,(t/t_{c})^{-3/5}\,{\rm km\,s^{-1}} &({\rm for}\,\, t\gtrsim t_{c}^{\prime}), \\
\end{array}\right. 
\end{equation}
where $V_{\rm free}=\sqrt{2E_{\rm out}/M_{\rm Bon}}$ is the free expansion velocity, 
$t_{\rm c}^{\prime}\approx 4.0\,$yr, and $R_{\rm c}^{\prime}\approx 3.9\times 10^{16}\,$cm from 
by setting the free expansion velocity equal to the velocity during the Sedov phase.
The thermal emission of the Bondi sphere dominates at this moment ($t_{\rm c}^{\prime}$), and ends 
the free expansion, and the optical depth of the Bondi sphere is  
$\tau_{\rm es}^{0}\approx 3\kappa_{\rm es}M_{\rm Bon}/4\pi {R_{\rm c}^{\prime}}^{2}=21.9$
for $M_{\rm Bon}=2.1\times 10^{2}\sunm$ and $R_{\rm c}^{\prime}=3.9\times 10^{16}\,{\rm cm}$. When the 
Bondi sphere becomes transparent, its thermal emission reaches the peak with the timescale of photon diffusion
\begin{equation}\label{Eq:trisep}
t_{\rm rise}^{\prime}=\frac{\tau_{\rm es}^{0}R_{\rm c}^{\prime}}{c}
            =0.92\,\left(\frac{M_{\rm Bon,2}}{2.1}\right)
            \left(\frac{R_{\rm c,16}^{\prime}}{3.9}\right)^{-1}\,{\rm yr},
\end{equation}
where $R_{\rm c,16}^{\prime}=R_{\rm c}^{\prime}/10^{16}\,{\rm cm}$, and the peak luminosity
\begin{equation}
L_{\rm Bon}\approx \frac{E_{\rm out}}{t_{\rm rise}^{\prime}}
           =3.5\times 10^{44}\,E_{52}\left(\frac{M_{\rm Bon,2}}{2.1}\right)^{-1}
             \left(\frac{R_{\rm c,16}^{\prime}}{3.9}\right).
\end{equation}
Considering that blackbody radiation increases with the surface area of the sphere, 
the rise light curve is $L_{\rm Bon}\propto R_{\rm exp}^{2}\propto t^{4/5}$.
Meanwhile, the thermal energy decays with the expansion kinetic energy,
as $L_{\rm Bon}\propto V_{\rm exp}^{2}\propto t^{-6/5}$.
We approximate the light curve as
\begin{equation}
L_{\rm obs}=\left\{\begin{array}{ll}
L_{\rm Bon}\left(t/t_{\rm rise}^{\prime}\right)^{4/5} & ({\textrm{for}}\,\, t\lesssim t_{\rm rise}^{\prime}),\\ 
     &   \\
L_{\rm Bon}\left(t/t_{\rm rise}^{\prime}\right)^{-6/5} & (\textrm{for}\,\, t\gtrsim t_{\rm rise}^{\prime}).\\
\end{array}\right.
\end{equation}
The characteristic temperature of the Bondi sphere peaks at
\begin{equation}
T_{\rm peak}=\left(\frac{L_{\rm Bon}}{4\pi\sigma_{\rm ST}{R_{\rm c}^{\prime}}^{2}}\right)^{1/4}
  \approx 4200\,\left(\frac{L_{\rm Bon,44}}{3.5}\right)^{1/4}
   \left(\frac{R_{\rm c,16}^{\prime}}{3.9}\right)^{-1/2}\,{\rm K},
\end{equation}
where $\sigma_{\rm ST}=5.67\times 10^{-5}\,{\rm erg\,s^{-1}\,cm^{-2}\,K^{-4}}$ is the Stefan-Boltzman 
constant, and $L_{\rm Bon,44}=L_{\rm Bon}/10^{44}\rm erg\,s^{-1}$. Moreover, the Bondi explosion may 
convert thermal energy into non-thermal energy through strong shocks when they pass through the BLR.

The Bondi sphere, as it expands with the Sedov velocity, generates shocks in the BLR.
We approximate the magnetic field of the post-shock gas as $B=1.9\,(n_{\rm BLR,7}T_{8})^{1/2}\,$G, where
$T_{8}=T/10^{8}\,$K. In this case, $u_{\rm B}\approx u_{\rm ph}$. With a radiative efficiency 
$\xi=0.1$, the non-thermal emission due to gas shocked by the Bondi explosion peaks at frequencies
\begin{equation}\label{Eq:nu}
\nu_{\rm syn}=10.5\,\left(\frac{B}{1.9\,\rm G}\right)\left(\frac{\gamma_{\rm max,5}}{5.7}\right)^{2}\,{\rm keV},\quad
\nu_{\rm IC}=1.4\,\left(\frac{\gamma_{\rm max,5}}{5.7}\right)^{2}\nu_{\rm UV,15}\,{\rm TeV},
\end{equation}
at a luminosity
\begin{equation}\label{Eq:L}
L_{\rm syn}^{\rm Bon}\approx 
L_{\rm IC}^{\rm Bon}\approx 1.7\times 10^{43}\,\xi_{0.1}E_{52}
           \left(t/t_{\rm rise}^{\prime}\right)^{-1}\,{\rm erg\,s^{-1}},
\end{equation}
where $\gamma_{\rm max,5}=\gamma_{\rm max}/10^{5}$.
The synchrotron radiation is approximately equal to the IC since $u_{\rm B}\approx u_{\rm ph}$. The 
peak luminosity in the X-rays may be marginally detectable when compared to the expected level of 
emission from the SMBH disk. Estimating the radio emission as a power-law ($L_{\nu}\propto \nu^{-0.5}$), 
$L_{5\rm GHz}\approx 10^{39}\,{\rm erg\,s^{-1}}$ at 5\,GHz, which is comparable to the brightness of 
AGNs of moderate radio power \citep[e.g.,][]{Elvis1994}. The rise and decay light curves
follow $t^{4/5}$ and $t^{-6/5}$, respectively, which can be used as a diagnostic of the event from
AGN light curves.

We note that \cite{Kimura2021} recently studied the evolution of BBHs formed in nuclear star 
clusters that are trapped by the SMBH disk. Their model depends on the in situ formation of BBHs in a 
nuclear star cluster, a scenario different from ours. We consider BBHs formed in the cavities of type~I AMS. 
The rejuvenation of AMS due to Bondi accretion of BBHs efficiently removes the orbital AM of the binaries,
leading to a merger event and a GW burst.  Meanwhile, a Bondi explosion driven by the powerful outflows 
from the hyper-Eddington accretion onto the BBH gives rise to an EMC.
Detailed numerical simulations have been done for the orbital evolution of BBHs in SMBH disks 
\citep[][]{Li2021}, but they do not include outflows. In this paper, we omit detailed discussions 
of the accretion onto BBHs, such as in the individual mini disks of each BH and the circumbinary disk
\citep{Kimura2021}. Our discussions should be valid since the timescale for forming the central cavity 
could be significantly longer than the characteristic timescales $t_{\rm a}$ and $t_{\rm vis}$. Moreover, 
the initial separation of the BBH is much smaller than the Bondi
radius. The behavior of Bondi–Hoyle–Lyttleton accretion onto BHHs \citep{Antoni2019,Comerford2019}
differs from the classical one \citep[e.g.,][]{Artymowicz1993}. A cavity around the BBH  
is in principle formed on a dynamical timescale when the disk around BBH is geometrically thin,
but the current case of hyper-Eddington accretion (geometrically thick) is more uncertain. 

The present scenario of BZ-powered emission and Bondi explosion could be a possible mechanism
to drive AGN variability on timescales of months to years. The event rates depend on the number of 
BHs. Investigating AGN light curves require continuous, long-term observational campaigns with 
suitable cadence in order to capture their rise and decay for comparison with theoretical predictions.
Radiation from the AMS depends on its BH mass, locations and the vertical structure of the SMBH disks.
In the work of \cite{Perna2021} and \cite{Zhu2021}, AMSs are located at $10^{3}\Rg$. Moreover,
they consider the influence of the vertical structure of the SMBH disks on AMSs of neutron stars. 
In future papers we will test how the properties of AMSs change with location within the SMBH disk.

\subsection{The case of GW\,190521}
We briefly apply the current model to explain the case of GW\,190521. This GW burst is a merger of
$85\sunm+66\sunm$ BHs, which are much higher than 
the upper limit of BHs produced by isolated massive stars driven by pair 
instability \citep[e.g.,][]{Woosley2002}. Our model of AMS in SMBH disks provides a promising 
framework for stellar-mass BHs to rapidly grow to exceed the pair instability limit. For an 
exponentially growing AMS BH, we have
$m_{\bullet}=m_{\bullet}^{0}\exp\left(\langle\dot{m}_{\bullet}\rangle t/t_{\rm Salp}\right)$, where
$m_{\bullet}^{0}$ is the initial mass of the BH,
$\langle\dot{m}_{\bullet}\rangle=\dot{m}_{\bullet}\delta_{\bullet}$ is the average rate of accretion 
onto the BH over time $t$, $\dot{m}_{\bullet}=f_{\rm a}\dot{m}_{\rm Bon}$ and 
$f_{\rm a}$ is the fraction of the accretion rate channeled into the outflow \citep{Takeo2020,Wang2021}, 
and $t_{\rm Salp}=\bhm/\dot{M}_{\rm Edd}=0.45\,$Gyr is the Salpeter time. The exponential growth reads
\begin{equation}\label{Eq:growth}
m_{\bullet}=m_{\bullet}^{0}\exp\left[6\,
           \left(\frac{\langle\dot{m}_{\bullet}\rangle_{6}t}{t_{\rm Salp}}\right)\right]
           \approx 7.4\,m_{\bullet}^{0},
\end{equation}
for $t=t_{\rm Salp}/3$, where 
$\langle\dot{m}_{\bullet}\rangle_{6}=(\delta_{\bar{6}}/6)\dot{m}_{9}f_{\bar{3}}$, 
$\delta_{\bar{6}}=\delta_{\bullet}/10^{-6}$ and $\dot{m}_{9}=\dot{M}_{\rm Bon}/10^{9}\dot{M}_{\rm Edd}$
for BHs with initial mass of $10\sunm$ \citep[see Eq. 20 in][]{Wang2021}.
Most of the uncertainty in the growth rate derives from $f_{\rm a}$, 
but it is easy for the BH to grow to $\sim10^{2}\sunm$ from $10\sunm$ within one AGN lifetime 
($t=t_{\rm AGN}$) in the context of SMBH disks. 

Three kinds of EMCs with GW bursts have been suggested in \S3.3. With
$m_{\bullet}\approx 150\sunm$, the characteristic of an EMC driven by the BZ-powered blob
is generally consistent with the flare of the quasar SDSS J1249+3449 monitored by the Zwicky 
Transient Facility \citep{Graham2020}. Moreover, there is a delay of $\sim 20$\,days (in the quasar's 
frame) of the EMC candidate (J1249+3449) with respect to the GW190521. This can be conveniently 
explained by $t_{\rm rise}$ in the case of an ejected blob. We thus prefer case A of the EMC of 
GW 190521 associated with BZ-powered ejected blob, as shown in Fig. 2. The predicted thermal 
emissions from the Bondi sphere and non-thermal emission from the
Bondi explosion can be tested observationally, as explored in an upcoming work, and all
calculations of radiations will be done in an self-consistent way.

\section{Discussion and Conclusions}
Compact objects (neutron stars and stellar-mass BHs) manifest themselves as accretion-modified 
stars (AMSs) in the accretion disks of supermassive BHs (SMBH disks) in AGNs. AMSs around 
stellar-mass BHs fall in two classes: cold gas-enshrouded type~I and hot gas-enshrouded type~II.  
This paper studies the dynamics of both classes in the context of SMBH disks. We show that most AMSs 
dynamically evolve into tight binary BHs (BBHs). The BBH evolves under the tidal torque of the 
central SMBH until it rejuvenates by hot gas accretion during the type~II phase. We show that 
the orbital angular momentum (AM)
of the BBH is efficiently removed by the gaseous drag during the initial type~I phase.
With an onset separation of $\sim 10^{2}\,r_{\rm g}$ in this phase, the subsequent release of GW 
radiation leads to the merger of the BBH after $\sim 3$ hour. The GW burst rate in a typical 
quasar is estimated to be $3.2\times 10^{-3}\,{\rm yr^{-1}}$ if we conservatively assume that 
the SMBH disk contains $N_{\bullet}=40$ BHs with $10^{2}\sunm$. The predicted GW frequency of 
$\sim 10^{2}\,$Hz is accessible by LIGO. Considering the entire quasar population at $z\lesssim 1$, 
the rate of GW bursts from AMS binaries is expected to be $\sim 3\,{\rm yr^{-1}}$. The binary 
formation rates given by Eq. (\ref{eq:Nbin}) are significantly higher than the tidal disruption 
rates of stars \citep[$\sim 10^{-4}\,{\rm yr^{-1}}$ depending on the SMBH mass in galactic 
centers, see details in][]{Wang2004}. However, GW rates depend on the number of BHs, 
which is currently poorly known. Since the orbit evolution is much shorter than the typical 
lifetime of AGNs, the binary formation rate can be traced by AGN flares.
Detection of $\sim 10^{2}\,$Hz GWs will reveal the content of stellar BHs in SMBH disks.  

Three kinds of electromagnetic counterparts (EMCs) of GW bursts are predicted, driven by
the Blandford-Znajek (BZ) mechanism or Bondi explosion from powerful outflows. BZ-powered EMCs
with mildly relativistic motion appear from radio to TeV bands with a rise time of a few $10^{6}$s
and decay with $t^{-6/5}$. Bondi spheres have thermal emission 
peaking in the optical, and the Bondi explosion driven by powerful outflows has detectable non-thermal
emissions from radio to $\gamma$-rays. The EMCs have a rise profile of $t^{4/5}$ and decay with
$t^{-6/5}$. Depending on the viscosity of the Bondi sphere, the EMCs could appear in three ways.
An efficient viscosity results in GW burst to lead the EMCs, but an inefficient viscosity 
postpones BBH to merge resulting in Bondi explosion leading as precursor of GW bursts. An moderate
viscosity may lead to simultaneous appearance of GW busts and EMCs but EMCs will last much longer
duration. Search for these flares from radio to 
$\gamma$-rays in AGNs and quasars using the Zwicky Transient 
Facility, {\it Swift} and {\it Fermi} observations will advance the understanding of 
the physics currently discussed. There is the potential to simultaneously detect GW bursts
corresponding to EMCs. 

We stress that the GW bursts should be redshifted or blueshifted if detected by LIGO because the 
BBH corotates with and merges within the SMBH disk.
The characteristic shifts would be $\sim \pm3000\,r_4^{-1/2}\kms$ depending on the direction of
motion of the BBH with respect to the observer and its distance to the SMBH.  These are pure Doppler 
shifts. Moreover, gravitational redshifts could also arise if BBH merges close enough to the 
SMBH \citep[e.g.,][]{Chen2019}. The Doppler and gravitational redshifts of GW bursts are important 
diagnostics to probe their birth place --- the SMBH disk. Lastly, we note that
the BBHs and the SMBH comprise another binary (with extreme mass ratio inspiral of $\sim 10^{-7}$)
and radiate GWs with frequencies of $f\sim 0.64\,M_{8}^{-1}r_{4}^{-3/2}$nano-Hz, which is only 
detectable through Pulsar Timing Arrays.

\acknowledgments
The authors are grateful to an anonymous referee for a useful report that clarified several 
points in this paper. Useful discussions are acknowledged with members from
IHEP AGN Group. We thank the support from the National Key R\&D Program of China (2016YFA0400701, 
2016YFA0400702, 2020YFC2201400), NSFC (NSFC-11991050, -11991054, -11833008, -11721303, -11991052, 
-11690024, QYZDJ-SSW-SLH007, XDB23010400), and by the International Partnership Program of Chinese 
Academy of Sciences (113111KYSB20200014).

\appendix
\section{BZ-powered jets from hyper-Eddington accretion}
The Blandford-Znajek mechanism is a powerful process of pumping energy from the BH spin
\citep{Blandford1977}. Given a BH with spin AM $\calJ_{\bullet}$, magnetic field $B$ normal 
to the horizon at $R_{\rm h}$, the pumping power is given by \citep{Ghosh1997,Macdonald1982}
\begin{equation}
L_{\rm BZ}=\left(\frac{1}{32}\right)\omega_{\rm F}^{2}B_{\perp}^{2}R_{\rm h}^{2}c
           \left(\frac{\calJ_{\bullet}}{\calJ_{\rm max}}\right)^{2},
\end{equation}
where $\calJ_{\rm max}$ is the maximum of the spin AM, 
$\omega_{\rm F}=\Omega_{\rm F}(\Omega_{\rm h}-\Omega_{\rm F})/\Omega_{\rm h}^{2}$ is the factor 
describing relative angular velocity of magnetic field to the BH ($\Omega_{\rm h}$).
BZ power has been calculated by \cite{Armitage1999} for optically thin advection-dominated accretion 
flows. Following this treatment, we use the 
self-similar solution of super-Eddington accretion \citep{Wang1999} for the energy channeled 
into a relativistic jet. We take the magnetic field of $B_{\perp}^{2}/8\pi=\alpha P_{\rm rad}$ 
(in equipartition with radiation field) and $R_{\rm h}=G\bhm/c^{2}$, where 
$P_{\rm rad}=\dot{M}_{\rm Bon}\Omega_{\rm K}/4\pi \alpha R$ is the radiation pressure of the 
super-Eddington accretion (dominated by gas pressure). 
Here we stress that the accretion rate of the BH $\dot{M}_{\bullet}=f_{\rm a}\dot{M}_{\rm Bon}$ 
is only a small fraction of the Bondi rate, 
and the factor $f_{\rm a}$ is uncertain \citep[see latest simulations of][]{Takeo2020}. 
For a maximally rotating BH, the BZ power
\begin{equation}
L_{\rm BZ}\approx \left(\frac{1}{64}\right)\dot{M}_{\bullet}c^{2}
          =2.0\times 10^{45}\,f_{\bar{3}}\dot{m}_{10}m_{2}\,{\rm erg\,s^{-1}},
\end{equation}
where $\omega_{\rm F}=1/2$ as taken by \cite{Armitage1999}, and the total pumped energy during 
the hyper-Eddington accretion episode ($t_{a}$) is given by
\begin{equation}
E_{\rm BZ}=L_{\rm BZ}t_{a}=2\times 10^{51}\,f_{\bar{3}}\dot{m}_{10}m_{2}t_{\rm a,6}\,{\rm erg},
\end{equation} 
where $f_{\bar{3}}=f_{\rm a}/10^{-3}$ and $t_{\rm a,6}=t_{\rm a}/10^{6}\,{\rm s}$.
\cite{Chen2021} recently expressed the BZ-powered jet in an approximate, analytical form. 
We take the outermost magnetic stream surface as the radius of the jet from their Eq.(105), 
\begin{equation}
R_{\rm jet}=C_{2}^{-1/2}\left(1+\sqrt{1-a^{2}}\right)^{\nu/2}\left(\frac{z}{r_{\rm g}}\right)^{1-\nu/2}
           \!\!\!r_{\rm g}
           =1.7\times 10^{12}\,R_{\rm Bon,15}^{5/8}m_{2}^{3/8}\,{\rm cm},
\end{equation}
where $C_{2}=\Gamma(3/2-\nu/2)\Gamma(1+\nu/2)/\sqrt{\pi}$ is a constant, $\Gamma$ is the $\Gamma$-function,
and $\nu$ is the power index of magnetic fields along the AMS radius. For the typical value of $\nu=3/4$,
$C_{2}=0.47$. The jet or the ejecta are choked by the dense medium of the 
SMBH disk \citep[e.g.,][]{Matzner2003,Zhu2021} to form a blob moving with a mildly relativistic velocity. 

\end{document}